\documentclass[12pt]{article}
\usepackage{a4wide,graphicx,float,cite}

\newcommand{\cD}{{\cal D}}
\newcommand{\cF}{{\cal F}}
\newcommand{\nn}{\nonumber}

\newcommand{\as}{\alpha_{s}}

\newcommand{\IM}{\mbox{\rm Im}}
\newcommand{\eqn}[1]{(\ref{#1})}
\newcommand{\mev}{\mbox{\rm MeV}}
\newcommand{\gev}{\mbox{\rm GeV}}

\newcommand{\sq}{\langle\bar ss\rangle}
\newcommand{\uu}{\langle\bar uu\rangle}

\newcommand{\qq}[1]{\langle\bar q_{#1}q_{#1}\rangle}

\newcommand{\newsection}[1]{\section{#1}\setcounter{equation}{0}}

\begin{document}

\begin{titlepage}
\phantom{XXX}
\vspace{-32mm}
\begin{flushright}
{\small\sf IFIC/02-61}\\[-1mm]
{\small\sf CAFPE/14-02}\\[-1mm]
{\small\sf UG-FT/144-02}\\[-1mm]
{\small\sf FTUV/02-1216}\\[-1mm]
{\small\sf HD-THEP-02-42}\\[8mm]
\end{flushright}

\begin{center}
{\Huge\bf\boldmath Determination of $m_s$ and $|V_{us}|$\\[3mm]
 from hadronic $\tau$ decays \unboldmath}\\[10mm]

{\normalsize\bf E.~G\'amiz${}^{1}$, M.~Jamin${}^{2,*}$, A.~Pich${}^{3}$,
 J.~Prades${}^{1}$ and F.~Schwab${}^{2}$} \\[4mm]

{\small\sl ${}^{1}$ Centro Andaluz de F\'{\i}sica de las Part\'{\i}culas
           Elementales (CAFPE) and}\\
{\small\sl Departamento de F\'{\i}sica Te\'orica y del Cosmos,
           Universidad de Granada,}\\
{\small\sl Campus de Fuente Nueva, E-18002 Granada, Spain}\\
{\small\sl ${}^{2}$ Institut f\"ur Theoretische Physik, Universit\"at
           Heidelberg,} \\
{\small\sl Philosophenweg 16, D-69120 Heidelberg, Germany}\\
{\small\sl ${}^{3}$ Departament de F\'{\i}sica Te\`orica, IFIC,
           Universitat de Val\`encia -- CSIC,}\\
{\small\sl Apt. Correus 22085, E-46071 Val\`encia, Spain} \\[10mm]
\end{center}

{\bf Abstract:} The mass of the strange quark is determined from SU(3)-breaking
effects in the $\tau$ hadronic width. Compared to previous analyses,
the contributions from scalar and pseudoscalar spectral functions, which
suffer from large perturbative corrections, are replaced by phenomenological
parametrisations. This leads to a sizeable reduction of the uncertainties in
the strange mass from $\tau$ decays. Nevertheless, the resulting $m_s$ value
is still rather sensitive to the moment of the invariant mass distribution
which is used for the determination, as well as the size of the quark-mixing
matrix element $|V_{us}|$. Imposing the unitarity fit for the CKM matrix, we
obtain $m_s(2\,\gev)=117\pm 17\,\mev$, whereas for the present Particle Data
Group average for $|V_{us}|$, we find $m_s(2\,\gev)=103\pm 17\,\mev$. On the
other hand, using an average of $m_s$ from other sources as an input, we are
able to calculate the quark-mixing matrix element $|V_{us}|$, and we
demonstrate that if the present measurement of the hadronic decay of the $\tau$
into strange particles is improved by a factor of two, the determination of
$|V_{us}|$ is more precise than the current world average.

\vfill


\vspace{4mm}
{\small ${}^{*}$ Heisenberg fellow.}
\end{titlepage}


\newsection{Introduction}

About a decade after the development of QCD sum rules by Shifman, Vainshtein
and Zakharov \cite{svz:79}, it was realised that the hadronic decay of the
$\tau$ lepton could serve as an ideal system to study low-energy QCD under
rather clean conditions \cite{bra:88,np:88,bra:89,bnp:92}. Shortly afterwards,
the experimental precision of $\tau$ decay data was sufficiently improved in
order to explore this possibility in practice. Since then, detailed
investigations of the $\tau$ hadronic width as well as invariant mass
distributions have served to determine the QCD coupling $\as$ to a precision
competitive with the current world average
\cite{aleph:93,cleo:95,aleph:98,opal:99}. More recently, the experimental
separation of the Cabibbo-allowed decays and Cabibbo-suppressed modes into
strange particles opened a means to also determine the mass of the strange
quark \cite{pp:98,ckp:98,aleph:99,pp:99,kkp:00,km:00,dchpp:01,cdghpp:01},
one of the fundamental QCD parameters within the Standard Model.

Until today, these strange mass determinations suffer from sizeable
uncertainties due to higher order perturbative corrections. In the sum rule
under investigation, scalar and pseudoscalar correlation functions contribute,
which are known to be inflicted with large higher order QCD corrections
\cite{bnp:92,ck:93,mal:98a,pp:98}, and these corrections are additionally
amplified by the particular weight functions which appear in the $\tau$ sum
rule. As a natural continuation, it was realised that one remedy of the
problem would be to replace the QCD expressions of scalar and pseudoscalar
correlators by corresponding phenomenological hadronic parametrisations
\cite{aleph:99,pp:99,km:00,mk:01}, which are expected to be more precise than
their QCD counterparts. In this work, we shall present a complete analysis of
this approach, and it will be shown that the determination of the strange quark
mass can indeed be significantly improved.

By far the dominant contributions to the pseudoscalar correlators come from
the kaon and the pion, which are very well known. The corresponding parameters
for the next two higher excited states have been recently estimated
\cite{mk:02}. Though much less precise, the corresponding contributions to
the $\tau$ sum rule are suppressed, and thus under good theoretical control.
The remaining strangeness-changing scalar spectral function has been extracted
very recently from a study of S-wave $K\pi$ scattering \cite{jop:00,jop:01} in
the framework of chiral perturbation theory ($\chi$PT) \cite{gl:84,gl:85} with
explicit inclusion of resonances \cite{egpr:89,eglpr:89}. The resulting scalar
spectral function was then employed to directly determine the strange quark
mass from a purely scalar QCD sum rule \cite{jop:02}. On the other hand, now
we are also in a position to incorporate this contribution into the $\tau$ sum
rule. The scalar $ud$ spectral function is still only very poorly determined
phenomenologically, but it is well suppressed by the small factor $(m_u-m_d)^2$
and can be safely neglected.

In the next section, for completeness, we begin by recalling the theoretical
expressions which are required in the sum rules under investigation. Then, in
section~3, the scalar and pseudoscalar contributions to the $\tau$ sum rules
are investigated separately, and it is demonstrated that, although the
uncertainties on the QCD side are large, they are well satisfied. In section~4,
an improved determination of the mass of the strange quark is performed by
replacing the QCD expressions for the scalar and pseudoscalar spectral
functions with the corresponding hadronic parametrisations. Since the resulting
strange quark mass depends quite sensitively on the value of the quark-mixing
matrix element $|V_{us}|$, in section~5 we take a different approach to the
considered sum rule.

Assuming a strange mass as extracted from other recent analyses, we use the
$\tau$ sum rule to determine $|V_{us}|$. The resulting uncertainty on
$|V_{us}|$ is still sizeable, but the error is dominated by the partial
hadronic width of the $\tau$ decaying into strange particles. It is then
demonstrated that an improvement of this measurement by a factor of two would
result in a determination of $|V_{us}|$ better than the present world average,
thus shedding some light on the question of the unitarity violation in the
first row of the Cabibbo-Kobayashi-Maskawa (CKM) matrix. Finally, in section~6
we compare our results with other recent determinations of the strange quark
mass and in section~7 we end with a discussion and conclusions.

\newsection{Theoretical framework}

Below, we review the main theoretical expressions required in the theoretical
analysis of the inclusive hadronic $\tau$ decay width. Further details and
complete expressions can be found in the original works
\cite{bnp:92,pp:98,pp:99}. The central quantities in such an analysis are
the two-point correlation functions
\begin{equation}
\label{PiVAmunu}
\Pi_{\mu\nu,ij}^{V/A}(p) \;\equiv\; i \int \! dx \, e^{ipx} \,
\langle\Omega| \, T\{ J_{\mu,ij}^{V/A}(x)\,J_{\nu,ij}^{V/A}(0)^\dagger\}|
\Omega\rangle\,,
\end{equation}
where $\Omega$ denotes the physical vacuum and the hadronic vector/axialvector
currents are given by
$J_{\mu,ij}^{V/A}(x)=(\bar q_j\gamma_\mu(\gamma_5)q_i)(x)$. The indices $i,j$
denote the light quark flavours up, down and strange. The correlators
$\Pi_{\mu\nu,ij}^{V/A}(p)$ have the Lorentz decomposition
\begin{equation}
\Pi_{\mu\nu,ij}^{V/A}(p) \;=\; (p_\mu p_\nu-g_{\mu\nu}p^2)\,\Pi^{V/A,T}_{ij}
(p^2) + p_\mu p_\nu\,\Pi^{V/A,L}_{ij}(p^2) \,,
\end{equation}
where the superscripts in the transversal and longitudinal components denote
the corresponding angular momentum $J=1$ ($T$) and $J=0$ ($L$) in the hadronic
rest frame.

The hadronic decay rate of the $\tau$ lepton,
\begin{equation}
\label{RTauex}
R_\tau \;\equiv\; \frac{\Gamma(\tau^-\to{\rm hadrons}\,\nu_\tau(\gamma))}
{\Gamma(\tau^-\to e^-\bar\nu_e\nu_\tau(\gamma))} \;=\;
R_{\tau,V} + R_{\tau,A} + R_{\tau,S} \,,
\end{equation}
can be expressed as an integral of the spectral functions $\IM\,\Pi^T(s)$ and
$\IM\,\Pi^L(s)$ over the invariant mass $s=p^2$ of the final state hadrons
\cite{tsai:71}:
\begin{equation}
\label{RTauth}
R_\tau \;=\; 12\pi\int\limits_0^{M_\tau^2} \frac{ds}{M_\tau^2}\,\biggl(
1-\frac{s}{M_\tau^2}\biggr)^2 \biggl[\,\biggl(1+2\frac{s}{M_\tau^2}\biggr)
\IM\,\Pi^T(s)+\IM\,\Pi^L(s)\,\biggr] \,.
\end{equation}
The appropriate combinations of the two-point correlation functions are given
by
\begin{equation}
\Pi^J(s) \;\equiv\; |V_{ud}|^2\Big[\,\Pi^{V,J}_{ud}(s) + \Pi^{A,J}_{ud}(s)\,
\Big] + |V_{us}|^2\Big[\,\Pi^{V,J}_{us}(s) + \Pi^{A,J}_{us}(s)\,\Big] \,,
\end{equation}
with $V_{ij}$ being the corresponding matrix elements of the
Cabibbo-Kobayashi-Maskawa quark-mixing matrix. As has been indicated in
eq.~\eqn{RTauex}, experimentally, one can disentangle vector from axialvector
contributions in the Cabibbo-allowed ($\bar ud$) sector, whereas such a
separation is problematic in the Cabibbo-suppressed ($\bar us$) sector, since
$G$-parity is not a good quantum number in modes with strange particles.

Additional information can be inferred from the measured invariant mass
distribution of the final state hadrons. The corresponding moments
$R_\tau^{kl}$, defined by \cite{dp:92b}
\begin{equation}
\label{Rtaukl}
R_\tau^{kl} \;\equiv\; \int\limits_0^{M_\tau^2} \, ds \,
\biggl( 1 -\frac{s}{M_\tau^2} \biggr)^k \biggl(\frac{s}{M_\tau^2}\biggr)^l\,
\frac{d R_\tau}{ds} \,,
\end{equation}
can be calculated theoretically in analogy to $R_\tau = R_\tau^{00}$.
Exploiting the analytic properties of $\Pi^J(s)$, the moments \eqn{Rtaukl}
can be expressed as contour integrals in the complex $s$-plane running
counter clockwise around the circle $|s|=M_\tau^2$:
\begin{equation}
\label{Rtauklth}
R_\tau^{kl} \;=\; -\,i\pi\!\oint\limits_{|x|=1} \frac{dx}{x}\,
\Big[\, 3\,\cF^{kl}_{L+T}(x)\,\cD^{L+T}(M_{\tau}^2 x) + 4\,\cF^{kl}_L(x)\,
\cD^L(M_{\tau}^2 x)\,\Big] \,.
\end{equation}
Here, we have defined the new integration variable $x\equiv s/M_\tau^2$, and
general expressions for the kinematical kernels $\cF^{kl}_{L+T}(x)$ and
$\cF^{kl}_L(x)$ have been presented in ref.~\cite{pp:99}. For the convenience
of the reader, we have compiled explicit expressions for the moments $(0,0)$
to $(4,0)$, which will be utilised in our phenomenological analysis in
table~\ref{tab1}.

\begin{table}[htb]
\renewcommand{\arraystretch}{1.1}
\begin{center}
\begin{tabular}{|c|c|c|}
\hline
$(k,l)$ & $\qquad\qquad\cF^{kl}_{L+T}(x)\qquad\qquad$ &
$\qquad\cF^{kl}_{L}(x)\qquad$
\\ \hline
(0,0) & $(1-x)^3\, (1+x)$ &  $(1-x)^3$  \\
(1,0) & $\frac{1}{10}\, (1-x)^4\, (7+8x)$ &  $\frac{3}{4}\, (1-x)^4$ \\
(2,0) & $\frac{2}{15}\, (1-x)^5\, (4+5x)$ &  $\frac{3}{5}\, (1-x)^5$ \\
(3,0) & $\frac{1}{7} \, (1-x)^6\, (3+4x)$ &  $\frac{1}{2}\, (1-x)^6$ \\
(4,0) & $\frac{1}{14}\, (1-x)^7\, (5+7x)$ &  $\frac{3}{7}\, (1-x)^7$ \\
\hline
\end{tabular}
\end{center}
\caption{Explicit expressions for the kinematical kernels $\cF^{kl}_{L+T}(x)$
and $\cF^{kl}_{L}(x)$ which will be used in our phenomenological analysis.
\label{tab1}}
\end{table}

In the derivation of the above expression \eqn{Rtauklth}, integration by parts
has been employed to rewrite $R_\tau^{kl}$ in terms of the physical correlators
$\cD^{L+T}(s)$ and $\cD^L(s)$,
\begin{equation}
\label{DLTDL}
\cD^{L+T}(s) \;\equiv\; -s\,\frac{d}{ds}\,\Big[\Pi^{L+T}(s)\Big] \,,
\qquad
\cD^{L}(s) \;\equiv\; \frac{s}{M_\tau^2}\,\frac{d}{ds}\,\Big[s\,\Pi^{L}(s)
\Big] \,,
\end{equation}
which both satisfy homogeneous renormalisation group equations, and thus
eliminate the dependence on unphysical (renormalisation scale and scheme
dependent) subtraction constants. For large enough negative $s$, the
contributions to $\cD^J(s)$ can be organised in the framework of the operator
product expansion (OPE) in a series of local gauge-invariant operators of
increasing dimension $D=2n$ times appropriate inverse powers of $s$. This
expansion is expected to be well behaved along the complex contour
$|s|=M_\tau^2$, except at the crossing point with the positive real axis
\cite{pqw:76}. As can be seen from eq.~\eqn{Rtauklth} and table~\ref{tab1},
however, the contribution near the physical cut at $s=M_\tau^2$ is strongly
suppressed by a zero of order three or larger. Therefore, uncertainties
associated with the use of the OPE near the time-like axis are expected to be
very small.

Inserting the OPE series for $\cD^J(s)$ into \eqn{Rtauklth} and performing
the contour integration, the resulting expression for $R_\tau^{kl}$ can be
written as \cite{bnp:92}
\begin{equation}
R_\tau^{kl} \;=\; 3\,(|V_{ud}|^2+|V_{us}|^2) S_{{\rm EW}}\biggl\{\, 1 +
\delta^{kl(0)} + \sum\limits_{D\geq2}\Big( \cos^2\!\theta_C\,
\delta_{ud}^{kl(D)} + \sin^2\!\theta_C\,\delta_{us}^{kl(D)} \Big)\biggr\}
\end{equation}
where $\sin^2\!\theta_C\equiv |V_{us}|^2/(|V_{ud}|^2+|V_{us}|^2)$, $\theta_C$
being the Cabibbo angle, and the electroweak radiative correction
$S_{{\rm EW}}=1.0201\pm 0.0003$ \cite{ms:88,bl:90,erl:02} has been pulled out
explicitly. Owing to chiral symmetry, the purely perturbative dimension-zero
contribution $\delta^{kl(0)}$ is identical for vector and axialvector parts.
The symbols $\delta_{ij}^{kl(D)}\equiv (\delta_{ud,V}^{kl(D)}+
\delta_{ud,A}^{kl(D)})/2$ stand for the average of vector and axialvector
contributions from dimension $D\geq 2$ operators which contain implicit
suppression factors $1/M_\tau^D$.

The separate measurement of Cabibbo-allowed and Cabibbo-suppressed decay
widths of the $\tau$ lepton \cite{aleph:99} allows one to pin down the flavour
SU(3)-breaking effects, dominantly induced by the strange quark mass. Defining
the difference
\begin{equation}
\label{delRtaukl}
\delta R_\tau^{kl} \;\equiv\; \frac{R_{\tau,V+A}^{kl}}{|V_{ud}^2|} -
\frac{R_{\tau,S}^{kl}}{|V_{us}^2|} \;=\;
3\,S_{{\rm EW}}\sum\limits_{D\geq 2}\Big(\delta_{ud}^{kl(D)} -
\delta_{us}^{kl(D)}\Big) \,,
\end{equation}
many theoretical uncertainties drop out since these observables vanish in the
SU(3) limit. In particular, they are free of possible flavour-independent
instanton as well as renormalon contributions which could mimic dimension-two
corrections.

As was already stated in the introduction, the longitudinal contributions
$\delta_{ij,L}^{kl(2)}$ are plagued with huge perturbative higher order
corrections, which in previous analyses resulted in large corresponding
uncertainties for the strange quark mass. Thus, in the following, we shall
replace the theoretical expressions for the longitudinal part by corresponding
phenomenological contributions, thereby strongly reducing those uncertainties
in the $m_s$ determination. Before we embark on the full investigation of the
moment differences $\delta R_\tau^{kl}$, to be able to judge the gain of the
intended replacements in the longitudinal sector, in the next section we shall
first investigate $\tau$-like sum rules for the longitudinal part alone.

\newsection{The longitudinal sector}

From eq.~\eqn{Rtauklth}, again making use of the analytic structure of the
correlation functions, one can deduce the following $\tau$-like sum rules for
the longitudinal contributions:
\begin{equation}
\label{RtauklL}
R_{ij,V/A}^{kl,L} \;=\; -\,24\pi^2 \int\limits_0^1 dx(1-x)^{2+k} x^{l+1}
\rho^{V/A,L}_{ij}(M_\tau^2 x) \;=\;
-\,4\pi i\!\oint\limits_{|x|=1}\frac{dx}{x}\,\cF^{kl}_L(x)\,
\cD^{V/A,L}_{ij}(M_{\tau}^2 x) \,,
\end{equation}
with $\rho^{V/A,L}_{ij}(s)\equiv\IM\,\Pi^{V/A,L}_{ij}(s)/\pi$ being the
longitudinal spectral functions. Note that for convenience, we have defined
$R_{ij,V/A}^{kl,L}$ without the appropriate CKM and electroweak correction
factors. The longitudinal correlators are directly related to corresponding
correlators $\Psi^{V/A}_{ij}(s)$ for the divergences of vector and axialvector
currents:
\begin{equation}
\label{PiL}
\Pi^{V/A,L}_{ij}(s) \;=\; \frac{1}{s^2}\,\Big[\,\Psi^{V/A}_{ij}(s)-
\Psi^{V/A}_{ij}(0)\,\Big] \,.
\end{equation}
Furthermore, via the equations of motion for the quark fields,
$\Psi^{V/A}_{ij}(s)$ is related to the two-point correlators
$\Pi^{S/P}_{ij}\!(s)$ for the scalar/pseudoscalar currents
$J_{ij}^{S/P}\!(x)=(\bar q_j(\gamma_5)q_i)(x)$, and the constants
$\Psi^{V/A}_{ij}(0)$ to the quark condensate from a Ward-identity
\cite{bro:81}:
\begin{equation}
\label{PsiPi}
\Psi^{V/A}_{ij}(s) \;=\; (m_i\mp m_j)^2\,\Pi^{S/P}_{ij}\!(s) \,,
\quad
\Psi^{V/A}_{ij}(0) \;=\; -\,(m_i\mp m_j)[\,\qq{i}\mp\qq{j}\,] \,,
\end{equation}
where $m_i$ and $m_j$ are the masses of the quark flavours $q_i$ and $q_j$
respectively, and $\qq{i}$ as well as $\qq{j}$ are the corresponding quark
condensates. The upper sign corresponds to the vector/scalar and the lower
sign to the axialvector/pseudoscalar case. For simplicity, the vacuum state
has been omitted in the condensates, and we would like to point out that the
bi-quark operators in eq.~\eqn{PsiPi} are {\em non}-normal-ordered
\cite{sc:88,jm:93}. The theoretical expressions for the longitudinal
correlators $\cD^L_{ij}(s)$ can either be obtained from
refs.~\cite{bnp:92,ck:93,pp:98,pp:99}, or can be calculated straightforwardly
from the second of eqs.~\eqn{DLTDL} together with eq.~\eqn{PiL} and the
expressions for $\Psi^{V/A}_{ij}(s)$ given in ref.~\cite{jm:95}. For further
details and the explicit expressions the reader is referred to these
references.

\begin{table}[thb]
\renewcommand{\arraystretch}{1.1}
\begin{center}
\begin{tabular}{|c|ccc|}
\hline
 & $R_{us,A}^{00,L}$ & $R_{us,V}^{00,L}$ & $R_{ud,A}^{00,L}$ \\
\hline
Theory: & $-0.144\pm0.024$ & $-0.028\pm0.021$ & $-(7.79\pm0.14)\cdot 10^{-3}$\\
Phenom: & $-0.135\pm0.003$ & $-0.028\pm0.004$ & $-(7.77\pm0.08)\cdot 10^{-3}$\\
\hline
\end{tabular}
\end{center}
\caption{Comparison of theoretical and phenomenological longitudinal
contributions to the (0,0) moment of the $\tau$ sum rule.
\label{tab2}}
\end{table}

Let us now proceed with a numerical investigation of the sum rules of
eq.~\eqn{RtauklL}. In table~\ref{tab2}, we display a comparison of theoretical
and phenomenological values of the longitudinal contributions
$R_{us,A}^{00,L}$, $R_{us,V}^{00,L}$ and $R_{ud,A}^{00,L}$ to the $\tau$ sum
rule. Due to the global mass squared factor, these results depend sensitively
on the quark masses. For definiteness, we have set the strange mass to
$m_s(2\,\gev)=105\,\mev$, a value compatible with most recent determinations
of $m_s$ both from QCD sum rules
\cite{mal:98,nar:99,km:00,kkp:00,cdghpp:01,mk:02,jop:02} and lattice QCD
\cite{lub:01,gm:01,kan:01,wit:02}. A detailed comparison to other $m_s$
determinations will be presented in section~6. The light masses $m_u$ and $m_d$
have been fixed from the $\chi$PT ratios $m_u/m_d=0.553\pm 0.043$ and
$m_s/m_d=18.9\pm 0.8$ \cite{leu:96}. On the theoretical side, we have also
included instanton contributions according to the model used in \cite{mk:02},
because they are known to play some r\^ole in scalar/pseudoscalar finite
energy sum rules.

The theoretical uncertainties in table~\ref{tab2} are largely dominated by
higher order perturbative corrections. Already for the $(0,0)$ moment, the
first and second order corrections are of similar size, while the third order
$\alpha_s^3$ correction is larger than these two. Thus for this moment, in
the spirit of asymptotic series, we have cut the perturbative series at the
second order. For higher moments, the behaviour is still worse and there the
second order turns out to be larger than the first one, which then entails
even larger perturbative uncertainties. For this reason, we have not presented
detailed results for the higher moments although also in these cases agreement
between theoretical and phenomenological results is found within the errors.
A detailed discussion of how we estimated the theoretical errors will be given
in the next section.

The uncertainties in the phenomenological numbers, on the other hand, are
much smaller. For the pseudoscalar spectral functions we have employed the
parametrisation by Maltman and Kambor \cite{mk:02}. In the case of the
pseudoscalar $us$ channel, the corresponding expression reads
\begin{equation}
\label{rhoSud}
s^2\rho^{A,L}_{us}(s) \;=\; 2f_K^2 M_K^4\delta(s-M_K^2) +
2f_1^2 M_1^4 B_1(s) + 2f_2^2 M_2^4 B_2(s) \,,
\end{equation}
with $f_i$ and $M_i$ being decay constants and masses of the higher exited
resonances. $B_i(s)$ is the standard Breit-Wigner resonance shape function
\begin{equation}
\label{BW}
B_i(s) \;=\; \frac{1}{\pi}\,\frac{\Gamma_i M_i}{[(s-M_i^2)^2+
\Gamma_i^2 M_i^2]} \,.
\end{equation}
For the pseudoscalar $ud$ channel an equation analogous to \eqn{rhoSud} holds.
The masses and widths of the resonances have been taken from the Review of
Particle Physics \cite{pdg:02} and the decay constants from ref.~\cite{mk:02}.
For convenience, we have compiled these input parameters in table~\ref{tab3}.
The dominant uncertainties in the phenomenological results of table~\ref{tab2}
are due to the errors in the decay constants $f_K$ and $f_\pi$, as well as the
variation of the higher resonance decay constants as given in table~\ref{tab3}.
Changing resonance masses and widths within reasonable ranges only has a minor
effect. Therefore, for these quantities, we have not given explicit
uncertainties.

\begin{table}[thb]
\renewcommand{\arraystretch}{1.1}
\begin{center}
\begin{tabular}{|r|cccc|}
\hline
 & $\pi(1300)$ & $\pi(1800)$ & $K(1460)$ & $K(1800)$ \\
\hline
$M_i\;[\mev]$ & $1300$ & $1800$ & $1460$ & $1830$ \\
$\Gamma_i\;[\mev]$ & $400$ & $210$ & $260$ & $250$ \\
$f_i\;[\mev]$ & $2.20\pm0.46$ & $0.19\pm0.19$ & $21.4\pm2.8$ & $4.5\pm4.5$ \\
\hline
\end{tabular}
\end{center}
\caption{Input parameters for the pseudoscalar resonance parametrisation. The
masses and widths are taken from \cite{pdg:02}, whereas the decay constants
have been determined in \cite{mk:02}.\label{tab3}}
\end{table}

The scalar $us$ spectral function has been calculated in ref.~\cite{jop:02}.
Below $2\,\gev$ the dominant hadronic systems which contribute in this channel
are the $K\pi$ and $K\eta'$ states with $K_0^*(1430)$ being the lowest lying
scalar resonance. The scalar $us$ spectral function can then be parametrised
in terms of the scalar $F_{K\pi}(s)$ and $F_{K\eta'}(s)$ form factors. These
form factors were obtained in \cite{jop:01} from a coupled-channel
dispersion-relation analysis. The S-wave $K\pi$ scattering amplitudes which
are required in the dispersion relations were available from a description
of S-wave $K\pi$ scattering data in the framework of unitarised $\chi$PT with
resonances \cite{jop:00}. The dominant uncertainty in the scalar $us$ spectral
function is due to an integration constant which emerges while solving the
coupled channel dispersion relations. In \cite{jop:01} it was decided to fix
this constant by demanding $F_{K\pi}(M_K^2-M_\pi^2)=1.22\pm 0.01$, a result
which from $\chi$PT is known to be very robust, since higher-order chiral
corrections are well suppressed \cite{gl:85}. The uncertainty in the
phenomenological value for $R_{us,V}^{00,L}$ in table~\ref{tab2} thus
corresponds to a variation of this parameter within the given range. Further
details on the scalar $us$ spectral function can be found in ref.~\cite{jop:02}.

Generally, from table~\ref{tab2} one observes that the longitudinal sum rules
are very well satisfied. This is partly due to the fact that the subtraction
constant $\Psi^{A}_{ud}(0)$ dominates the $ud$ sum rule and also in the
pseudoscalar $us$ sum rule $\Psi^{A}_{us}(0)$ gives a large contribution.
In the leading order in $\chi$PT, these subtraction constants are fixed by
the pion and kaon pole contributions which led to the famous
Gell-Mann-Oakes-Renner (GMOR) relations \cite{gmor:68}. Therefore, we have a
well known piece which appears on both sides of the sum rule. Nevertheless, the
agreement within errors is non-trivial and also points to the fact that the
used light quark mass values are reasonable. However, the complete agreement in
the scalar $us$ case should be considered as accidental. From table~\ref{tab2}
it becomes obvious that the theoretical uncertainties in the flavour-breaking
$\tau$ sum rule can be greatly improved by replacing the theoretical
longitudinal contributions by the corresponding phenomenological ones.

The observed sensitivity of the longitudinal sum rules on the subtraction
constants allows to obtain further information on these quantities. For this
purpose it is convenient to rewrite the subtraction constants as
\begin{equation}
\label{gmor}
\Psi^{A}_{ud}(0) \;=\; 2f_\pi^2 M_\pi^2\,(\,1-\delta_\pi\,) \,,
\qquad
\Psi^{A}_{us}(0) \;=\; 2f_K^2 M_K^2\,(\,1-\delta_K\,) \,,
\end{equation}
where the leading term corresponds to the GMOR relations, and to actually
determine the higher order chiral corrections $\delta_\pi$ and $\delta_K$
\cite{gl:85,jam:02}. In addition, from the scalar subtraction constant
$\Psi^{V}_{us}(0)$, the flavour breaking ratio of the quark condensates
$\sq/\uu$ can be estimated. Now we have to also vary the strange quark mass,
and we chose $m_s(2\,\gev)=105\pm 20\,\mev$, the central value already
advocated above with an uncertainty which is slightly more generous than the
ranges obtained in the recent scalar and pseudoscalar sum rule determinations
\cite{mk:02,jop:02}.

Again varying all other input parameters according to the ranges presented
here and in the next section, we obtain:
\begin{equation}
\label{delPdelK}
\delta_\pi \;=\; 0.049 \pm 0.030 \,,
\qquad
\delta_K \;=\; 0.50 \pm 0.23 \,,
\qquad
\frac{\sq}{\uu} \;=\; 0.80 \pm 0.58 \,.
\end{equation}
These results are in very good agreement to the findings of ref.~\cite{jam:02},
where the values of $\delta_\pi=0.047\pm 0.017$ and $\delta_K=0.61\pm 0.22$
were estimated in the framework of $\chi$PT assuming a given input for the
ratio $\sq/\uu$. However, the latter results of \cite{jam:02} are somewhat
more precise, since the uncertainties in eq.~\eqn{delPdelK} are again dominated
by the large higher order perturbative corrections and now in addition
by the error on the strange quark mass. Besides from the scalar subtraction
constant, the ratio $\sq/\uu$ can also be directly inferred from $\delta_K$,
and from $\delta_\pi$ by inverting the line of reasoning applied in
ref.~\cite{jam:02}. The central values found in this way are $\sq/\uu=0.63$
and $0.75$ respectively, however, with uncertainties which are even larger
than the one presented in eq.~\eqn{delPdelK}. Nevertheless, it is gratifying
to observe that the central results for $\sq/\uu$ display good consistency.

\newsection{Strange quark mass}

Our determination of the strange quark mass from the SU(3)-breaking differences
of eq.~\eqn{delRtaukl} proceeds in complete analogy to the previous analyses
\cite{pp:99,cdghpp:01}, with the exception that now we have replaced the
longitudinal contributions to the sum rule by phenomenological parametrisations
as discussed above. Solving for the strange mass, eq.~\eqn{delRtaukl} can be
written as
\begin{equation}
\label{msmtau}
m_s^2(M_\tau) \;=\; \frac{M_\tau^2}{18(1-\epsilon_d^2)\Delta_{kl}^{L+T}
(a_\tau)}\Biggl\{\,\frac{\delta R_\tau^{kl}}{S_{{\rm EW}}} -
\delta R^{kl,L}_{\tau,{\rm phen}} - \delta R^{kl,L+T}_{\tau,D\geq4}\,\Biggr\}\,,
\end{equation}
where $\epsilon_d\equiv m_d/m_s$. The phenomenological contribution containing
the longitudinal spectral functions is given by
\begin{equation}
\label{delRphen}
\delta R^{kl,L}_{\tau,{\rm phen}} \;=\;
R_{ud,V}^{kl,L} + R_{ud,A}^{kl,L} - R_{us,V}^{kl,L} - R_{us,A}^{kl,L} \,,
\end{equation}
with $R_{ij,V/A}^{kl,L}$ as defined in eq.~\eqn{RtauklL} and calculated using
the scalar and pseudoscalar spectral functions discussed in the last section.

The perturbative QCD correction $\Delta_{kl}^{L+T}(a_\tau)$, associated with
the dimension-2 contribution, depends only on $a_\tau\equiv\as(M_\tau)/\pi$
and has been discussed in great detail in refs.~\cite{pp:98,pp:99}. Since we
have subtracted the badly behaved longitudinal contribution, the remaining
part displays a better convergence at higher orders. In contour-improved
perturbation theory \cite{dp:92a}, and taking as a central value
$\as(M_\tau)=0.334$ \cite{aleph:98}, the expansion for the relevant moments
takes the form
\begin{eqnarray}
\label{DeltaLT}
\Delta_{00}^{L+T}(a_\tau) \!&=&\! 0.753 + 0.214 + 0.065 - 0.051 + \ldots
\;=\; 0.981 \,, \nn \\
\Delta_{10}^{L+T}(a_\tau) \!&=&\! 0.912 + 0.334 + 0.192 + 0.056 + \ldots
\;=\; 1.493 \,, \nn \\
\Delta_{20}^{L+T}(a_\tau) \!&=&\! 1.055 + 0.451 + 0.330 + 0.189 + \ldots
\;=\; 2.024 \,, \\
\Delta_{30}^{L+T}(a_\tau) \!&=&\! 1.190 + 0.571 + 0.484 + 0.352 + \ldots
\;=\; 2.597 \,, \nn \\
\Delta_{40}^{L+T}(a_\tau) \!&=&\! 1.324 + 0.697 + 0.657 + 0.551 + \ldots
\;=\; 3.228 \,. \nn
\end{eqnarray}
The last term in the series, corresponding to the order $\as^3$ correction,
has not yet been calculated. In eq.~\eqn{DeltaLT}, we have employed an estimate
based on assuming a geometrical growth of the Adler function $\cD^{L+T}(s)$
\cite{pp:99}. Finally, $\delta R^{kl,L+T}_{\tau,D\geq4}$ are higher-dimensional
operator corrections in the framework of the operator product expansion, the
dominant one being the $D=4$ correction resulting from the quark condensate.
Explicit expressions for these contributions can be found in ref.~\cite{pp:99}.
In our numerical analysis below, the $D=4$ light up and down quark mass
corrections as well as the $D=6$ condensates have also been included, although
their influence on the value of $m_s$ is insignificant.

\begin{table}[thb]
\renewcommand{\arraystretch}{1.1}
\begin{center}
\begin{tabular}{ccrrrrr}
\hline
Parameter & Value & $\quad(0,0)$ & $\quad(1,0)$ & $\quad(2,0)$ &
$\quad(3,0)$ & $\quad(4,0)$ \\
\hline
$m_s(M_\tau)$ & & 197.3 & 164.0 & 136.6 & 115.2 & 98.6 \\
\hline
$\delta R_\tau^{kl}$ & \cite{cdghpp:01} &
${}^{+52.7}_{-67.7}$ & ${}^{+22.2}_{-25.4}$ & ${}^{+13.1}_{-14.5}$ &
${}^{\phantom{1}+9.3}_{-10.2}$ & ${}^{+7.3}_{-7.9}$ \\
$|V_{us}|$ & $0.2225\pm0.0021$ &
${}^{+28.8}_{-32.6}$ & ${}^{+14.6}_{-16.0}$ & ${}^{\phantom{1}+9.4}_{-10.1}$ &
${}^{+6.7}_{-7.1}$ & ${}^{+5.3}_{-5.6}$ \\
${\cal O}(\as^3)$ & ${}^{2\times{\cal O}(\as^3)}_{{\rm no}\;{\cal O}(\as^3)}$ &
${}^{+5.9}_{-5.4}$ & ${}^{-3.1}_{+3.3}$ & ${}^{-6.0}_{+6.9}$ &
${}^{-7.1}_{+8.6}$ & ${}^{-7.5}_{+9.6}$ \\
$\xi$ & ${}^{1.5}_{0.75}$ &
${}^{-11.7}_{+29.0}$ & ${}^{\phantom{1}-1.5}_{+11.8}$ & ${}^{+3.5}_{+4.2}$ &
${}^{+6.3}_{-0.1}$ & ${}^{+7.9}_{-2.7}$ \\
$\as(M_\tau)$ & $0.334\pm0.022$ &
${}^{+14.2}_{\phantom{1}-9.8}$ & ${}^{+5.0}_{-3.0}$ & ${}^{+1.1}_{+0.1}$ &
${}^{-0.9}_{+1.8}$ & ${}^{-2.0}_{+2.8}$ \\
$\langle\bar ss\rangle/\langle\bar uu\rangle$ & $0.8\pm0.2$ &
${}^{-0.8}_{+0.8}$ & ${}^{-3.6}_{+3.5}$ & ${}^{-5.6}_{+5.3}$ &
${}^{-7.2}_{+6.8}$ & ${}^{-8.4}_{+7.9}$ \\
$\rho_{us}^{\rm scalar}$ & see text &
${}^{-2.0}_{+1.8}$ & ${}^{-0.8}_{+0.7}$ & ${}^{-0.4}_{+0.3}$ &
${}^{-0.2}_{+0.2}$ & ${}^{-0.1}_{+0.1}$ \\
$f_K$ & $113\pm1\,\mev$ &
${}^{-1.1}_{+1.1}$ & ${}^{-0.8}_{+0.8}$ & ${}^{-0.6}_{+0.6}$ &
${}^{-0.5}_{+0.5}$ & ${}^{-0.5}_{+0.4}$ \\
$f_{K(1460)}$ & $21.4\pm 2.8\,\mev$ &
${}^{-1.1}_{+0.9}$ & ${}^{-0.4}_{+0.4}$ & ${}^{-0.2}_{+0.2}$ &
${}^{-0.1}_{+0.1}$ & ${}^{-0.1}_{+0.1}$ \\
\hline
Total & &
${}^{+68.4}_{-76.9}$ & ${}^{+29.9}_{-30.6}$ & ${}^{+18.9}_{-19.5}$ &
${}^{+17.2}_{-16.0}$ & ${}^{+17.5}_{-15.2}$ \\
\hline
\end{tabular}
\end{center}
\caption{Central results for $m_s(M_\tau)$ corresponding to the unitarity fit
for $|V_{us}|$ extracted from the different moments, as well as ranges for the
main input parameters and resulting uncertainties for $m_s$. For a detailed
description see the discussion in the text.
\label{tab4}}
\end{table}

Due to cancellations in the difference \eqn{delRtaukl}, it was already found
in the earlier analyses \cite{pp:99,cdghpp:01,mk:02} that the strange mass
resulting from eq.~\eqn{msmtau} depends sensitively on the value for $|V_{us}|$.
Therefore, in what follows we shall perform two separate extractions of $m_s$.
In the first one, like in ref.~\cite{cdghpp:01}, the value for $|V_{us}|$ is
taken from the Particle Data Group unitarity fit of the CKM matrix \cite{pdg:02}
yielding $|V_{us}|=0.2225\pm 0.0021$. In table~\ref{tab4}, we show a detailed
account of our results. The first row displays the values of $m_s(M_\tau)$
obtained from the different moment sum rules and central values for all input
parameters. In the following rows, we have listed those input parameters which
dominantly contribute to the uncertainty on $m_s$, the ranges for these
parameters used in our analysis and the resulting shift in $m_s$. Only those
parameters have been included in the list which at least for one moment yield
a shift of $m_s$ larger than $1\,\mev$. Finally, in the last row, we display
the total error that results from adding the individual uncertainties in
quadrature.

Let us now discuss the different uncertainties in more detail. The experimental
results for $\delta R_\tau^{k0}$ with $k=0\ldots 4$, as well as the errors
corresponding to a variation of $|V_{us}|$, have been presented in ref.
\cite{cdghpp:01} and are taken over from there. They dominate for low $k$ and
become smaller if $k$ is increased. For the uncertainties related to higher
order perturbative corrections we have considered three different sources.
As already mentioned above, the order $\as^3$ correction to
$\Delta_{kl}^{L+T}(a_\tau)$ has not yet been calculated and thus we rely on
estimates of this correction \cite{bnp:92,dp:92a}. To be conservative, for our
error estimate we have chosen two extreme cases, one where this correction
is doubled and the other where it is removed completely. The corresponding
variations of $m_s$ are given in table~\ref{tab4}. The second source is the
renormalisation scale dependence which arises due to the missing higher orders
which would make physical quantities independent of the scale. Like in
refs.~\cite{dp:92a}, we have parametrised this dependence by the parameter
$\xi\equiv\mu/\sqrt{-s}$, and we have varied this parameter in the range
$0.75<\xi<1.5$. The upper range is slightly reduced compared to the previous
analyses, but since we have included explicitly the variation of the
${\cal O}(\as^3)$ correction, we consider this reduction justified. Finally,
our input value for $\as(M_\tau)=0.334\pm0.022$ \cite{cdghpp:01} reflects
itself in an additional uncertainty for the $m_s$ determination.

The only parameter which matters for the uncertainty related with the $D=4$
contribution in the OPE is the flavour-breaking ratio of the quark condensates
$\sq/\uu$. For this parameter we have chosen the range $\sq/\uu=0.8\pm0.2$
which includes most values found in the literature
\cite{djn:89,nar:89,ls:91,jm:95,drs:01}. Certainly, this value is also
compatible with the results obtained in section~3. For increasing $k$, higher
order terms in the OPE become more important, and thus for the larger $k$
values this error, together with the perturbative QCD uncertainty, dominates.
Although we have included the $D=6$ contribution according to the expressions
given in \cite{pp:99} in our numerical analysis, for the resulting value of
$m_s$ and thus also for the uncertainty they can be neglected.

The final uncertainties included in table~\ref{tab4} concern the variation of
the phenomenological scalar and pseudoscalar spectral functions. The variations
of the corresponding parameters have already been discussed in section~3,
where we have analysed $\tau$-like sum rules for the longitudinal contributions
separately. The dominant sources of error are the scalar $us$ spectral
function as discussed briefly in section~3, and in more detail in
ref.~\cite{jop:02}, as well as the decay constants in the pseudoscalar $us$
channel $f_K$ and $f_K(1460)$. However, as can be seen from table~\ref{tab4},
the influence on the error of the strange quark mass is very minor. In
addition, it again corroborates the reduction of the final uncertainty of
$m_s$ achieved through the replacement of theoretical by phenomenological
scalar and pseudoscalar longitudinal contributions. Variations of all remaining
input parameters lead to changes in $m_s$ of less than $1\,\mev$, and have
thus been neglected.

To calculate a final result for $m_s(M_\tau)$, we have used a weighted
average over the different moments where as the individual error for each
moment, we have taken the larger one. This prescription then yields:
\begin{equation}
\label{msUT}
m_s(M_\tau) \;=\; 121.7 \pm 17.2 \,\mev
\quad\Rightarrow\quad
m_s(2\,\gev) \;=\; 117 \pm 17 \,\mev \,.
\end{equation}
The result on the right-hand side is the value of $m_s$ evolved to a
renormalisation scale of $2\,\gev$. As the uncertainty we have taken the
smallest error of one individual moment, (3,0) in this case. Taking the
correlation of the different moments into account, the final uncertainty
could still be slightly reduced. However, since the moments are strongly
correlated, this would give little improvement, and thus at present we have
not incorporated this approach.

Comparing our central strange mass average to the $m_s$ values obtained for 
the individual moments, it is evident that there is a strong dependence of
$m_s$ on the number of the moment $k$. For our approach, this dependence is
even stronger than in the analysis of ref.~\cite{cdghpp:01}. The reason lies
in the fact that the scalar and pseudoscalar sum rules by themselves are
satisfied for smaller strange masses than the full $\tau$ sum rule. Thus,
when replacing the theoretical scalar and pseudoscalar spectral functions by
phenomenological ones, the resulting $m_s$ moves up. Nevertheless, one should
clearly state that the experimental uncertainties, especially for the low
moments are still large, and the difference between the individual moment
results and the average strange mass for all values lies in the range of one
standard deviation.

As was already mentioned above, the central PDG average for
$|V_{us}|=0.2196\pm 0.0026$ \cite{pdg:02}, which is based on the analyses
\cite{lr:84,cknrt:01,cl:02}, lies more than one sigma below the value
extracted by requiring unitarity for the first row of the CKM matrix. Due
to this large difference, we consider it justified to repeat our strange mass
determination for this value of $|V_{us}|$ separately. To properly perform
the respective analysis, one would in principle need the experimental results
for the $R$-ratios $R_{\tau,V+A}^{kl}$ and $R_{\tau,S}^{kl}$ for all moments
separately. However, as yet, these are only published for the lowest $(0,0)$
moment, whereas for the higher moments the combined result $\delta R_\tau^{kl}$
based on the unitarity value for $|V_{us}|$ is solely available
\cite{cdghpp:01}. Since in ref.~\cite{cdghpp:01} the uncertainty from varying
$|V_{us}|=0.2225\pm 0.0021$ was listed separately, we can nevertheless work
around this problem by rescaling $\delta R_\tau^{kl}$ appropriately to the
above PDG average.

\begin{table}[htb]
\renewcommand{\arraystretch}{1.1}
\begin{center}
\begin{tabular}{ccrrrrr}
\hline
Parameter & Value & $\quad(0,0)$ & $\quad(1,0)$ & $\quad(2,0)$ &
$\quad(3,0)$ & $\quad(4,0)$ \\
\hline
$m_s(M_\tau)$ & & 150.8 & 141.5 & 122.5 & 105.3 & 90.8 \\
\hline
$\delta R_\tau^{kl}$ & experiment &
${}^{\phantom{1}+61.9}_{-103.3}$ & ${}^{+25.0}_{-30.1}$ & ${}^{+14.5}_{-16.4}$&
${}^{+10.2}_{-11.3}$ & ${}^{+7.9}_{-8.7}$ \\
$|V_{us}|$ & $0.2196\pm0.0026$ &
${}^{+42.1}_{-56.6}$ & ${}^{+20.3}_{-23.5}$ & ${}^{+12.7}_{-14.2}$ &
${}^{+8.9}_{-9.8}$ & ${}^{+7.0}_{-7.6}$ \\
${\cal O}(\as^3)$ & ${}^{2\times{\cal O}(\as^3)}_{{\rm no}\;{\cal O}(\as^3)}$ &
${}^{+4.3}_{-3.9}$ & ${}^{-2.6}_{+2.8}$ & ${}^{-5.3}_{+6.1}$ &
${}^{-6.5}_{+7.9}$ & ${}^{-6.9}_{+8.9}$ \\
$\xi$ & ${}^{1.5}_{0.75}$ &
${}^{\phantom{1}-8.1}_{+20.2}$ & ${}^{-1.1}_{+9.7}$ & ${}^{+3.2}_{+3.6}$ &
${}^{+5.8}_{-0.1}$ & ${}^{+7.3}_{-2.5}$ \\
$\as(M_\tau)$ & $0.334\pm0.022$ &
${}^{+10.1}_{\phantom{1}-7.0}$ & ${}^{+4.2}_{-2.5}$ & ${}^{+1.0}_{+0.1}$ &
${}^{-0.8}_{+1.6}$ & ${}^{-1.9}_{+2.5}$ \\
$\langle\bar ss\rangle/\langle\bar uu\rangle$ & $0.8\pm0.2$ &
${}^{-1.2}_{+1.1}$ & ${}^{-4.3}_{+4.1}$ & ${}^{-6.3}_{+5.9}$ &
${}^{-8.0}_{+7.4}$ & ${}^{-9.2}_{+8.5}$ \\
$\rho_{us}^{\rm scalar}$ & see text &
${}^{-2.5}_{+2.2}$ & ${}^{-0.9}_{+0.8}$ & ${}^{-0.4}_{+0.4}$ &
${}^{-0.2}_{+0.2}$ & ${}^{-0.2}_{+0.1}$ \\
$f_K$ & $113\pm 1\,\mev$ &
${}^{-1.4}_{+1.4}$ & ${}^{-0.9}_{+0.9}$ & ${}^{-0.7}_{+0.7}$ &
${}^{-0.6}_{+0.6}$ & ${}^{-0.5}_{+0.5}$ \\
$f_{K(1460)}$ & $21.4\pm 2.8\,\mev$ &
${}^{-1.3}_{+1.2}$ & ${}^{-0.5}_{+0.4}$ & ${}^{-0.2}_{+0.2}$ &
${}^{-0.1}_{+0.1}$ & ${}^{-0.1}_{+0.1}$ \\
\hline
Total & &
${}^{\phantom{1}+78.4}_{-118.4}$ & ${}^{+34.2}_{-38.6}$ & ${}^{+21.4}_{-23.2}$&
${}^{+18.3}_{-18.1}$ & ${}^{+17.9}_{-16.6}$ \\
\hline
\end{tabular}
\end{center}
\caption{Central results for $m_s(M_\tau)$ corresponding to the PDG average
for $|V_{us}|$ extracted from the different moments, as well as ranges for the
main input parameters and resulting uncertainties for $m_s$. For a detailed
description see the discussion in the text.
\label{tab5}}
\end{table}

The result of our second analysis is displayed in table~\ref{tab5}. Like in
table~\ref{tab4}, again the first row contains the individual moment results
for central input parameters, then the variations of $m_s$ while varying the
dominant input parameters are tabulated, and the last row gives the combined
uncertainty by adding all errors in quadrature. The variation of the various
input parameters proceeds in complete analogy to the unitarity case discussed
above. Again performing a weighted average of the individual moment results
treating the uncertainties like for the unitarity $|V_{us}|$, the resulting
strange quark mass turns out to be
\begin{equation}
\label{msLR}
m_s(M_\tau) \;=\; 107.0 \pm 17.9 \,\mev
\quad\Rightarrow\quad
m_s(2\,\gev) \;=\; 103 \pm 17 \,\mev \,.
\end{equation}
We observe that for the lower value of $|V_{us}|$, we also find $m_s$ to be
lowered by roughly one sigma, but nevertheless certainly compatible with the
result of eq.~\eqn{msUT}. As table~\ref{tab5} shows, here the $k$ dependence
of $m_s$ is weaker than for the unitarity case, the reason being that now
the central $m_s$ lies much closer to the value preferred by the scalar and
pseudoscalar sum rules. Then the replacement of theoretical by phenomenological
spectral functions only shifts the resulting $m_s$ by a smaller amount.

\newsection{A novel route to \boldmath $|V_{us}|$ \unboldmath}

Since the quantities $\delta R_\tau^{kl}$ of eq.~\eqn{delRtaukl} are strongly
dependent on $|V_{us}|$, especially for low $k$, now we want to turn the table
and investigate the potential of determining $|V_{us}|$ from the SU(3)-breaking
$\tau$ sum rule. For this novel approach, we require a value for the strange
mass from other sources as an input so that we are in a position to calculate
$\delta R_\tau^{kl}$ from theory. Like in section~3, we have used
$m_s(2\,\gev)=105\pm 20\,\mev$ which is compatible with the recent sum rule
determinations \cite{mal:98,nar:99,mk:02,jop:02} and also with the lattice
results \cite{lub:01,gm:01,kan:01,wit:02}.\footnote{Further references to
original works can be found in these reviews.}

\begin{table}[thb]
\renewcommand{\arraystretch}{1.1}
\begin{center}
\begin{tabular}{ccccccc}
\hline
Parameter & Value & (0,0) & (1,0) & (2,0) & (3,0) & (4,0) \\
\hline
$\delta R_{\tau,th}^{kl}$ & central & $\;0.229\;$ & $\;0.270\;$ & $\;0.320\;$ &
$\;0.375\;$ & $\;0.436\;$ \\
\hline
$m_s(2\,\gev)$ & $105 \pm 20\,\mev$ &
${}^{+0.028}_{-0.023}$ & ${}^{+0.043}_{-0.036}$ & ${}^{+0.058}_{-0.048}$ &
${}^{+0.075}_{-0.062}$ & ${}^{+0.094}_{-0.077}$ \\
${\cal O}(\as^3)$ & ${}^{2\times{\cal O}(\as^3)}_{{\rm no}\;{\cal O}(\as^3)}$ &
${}^{-0.004}_{+0.004}$ & ${}^{+0.004}_{-0.004}$ & ${}^{+0.013}_{-0.013}$ &
${}^{+0.024}_{-0.024}$ & ${}^{+0.038}_{-0.038}$ \\
$\xi$ & ${}^{1.5}_{0.75}$ &
${}^{+0.007}_{-0.013}$ & ${}^{+0.001}_{-0.012}$ & ${}^{-0.007}_{-0.007}$ &
${}^{-0.018}_{+0.000}$ & ${}^{-0.032}_{+0.013}$ \\
$\as(M_\tau)$ & $0.334\pm0.022$ &
${}^{-0.007}_{+0.006}$ & ${}^{-0.005}_{+0.003}$ & ${}^{-0.001}_{-0.001}$ &
${}^{+0.004}_{-0.007}$ & ${}^{+0.011}_{-0.014}$ \\
$\langle\bar ss\rangle/\langle\bar uu\rangle$ & $0.8\pm0.2$ &
${}^{+0.002}_{-0.002}$ & ${}^{+0.010}_{-0.010}$ & ${}^{+0.018}_{-0.018}$ &
${}^{+0.024}_{-0.024}$ & ${}^{+0.030}_{-0.031}$ \\
$\rho_{us}^{\rm scalar}$ & see text &
${}^{+0.004}_{-0.004}$ & ${}^{+0.002}_{-0.002}$ & ${}^{+0.001}_{-0.001}$ &
${}^{+0.001}_{-0.001}$ & ${}^{+0.001}_{-0.001}$ \\
$f_K$ & $113\pm 1\,\mev$ &
${}^{+0.002}_{-0.002}$ & ${}^{+0.002}_{-0.002}$ & ${}^{+0.002}_{-0.002}$ &
${}^{+0.002}_{-0.002}$ & ${}^{+0.002}_{-0.002}$ \\
$f_{K(1460)}$ & $21.4\pm 2.8\,\mev$ &
${}^{+0.002}_{-0.002}$ & ${}^{+0.001}_{-0.001}$ & ${}^{+0.001}_{-0.001}$ &
${}^{+0.000}_{-0.000}$ & ${}^{+0.000}_{-0.000}$ \\
\hline
Total & error &
${}^{+0.030}_{-0.028}$ & ${}^{+0.044}_{-0.040}$ & ${}^{+0.062}_{-0.054}$ &
${}^{+0.083}_{-0.074}$ & ${}^{+0.107}_{-0.098}$ \\
\hline
\end{tabular}
\end{center}
\caption{Central results for $R_{\tau}^{kl}$ extracted from theory, as well as
ranges for the main input parameters and corresponding resulting uncertainties.
\label{tab6}}
\end{table}
Our results for the theoretical prediction of $\delta R_{\tau}^{kl}$ are
presented in table~\ref{tab6}. Like our results for the strange mass, the
first row shows our central values for $\delta R_{\tau}^{kl}$ and below
the uncertainties resulting from varying the various input parameters
within their ranges, now including the strange mass, are listed. Again, the
last row gives the total uncertainty by adding all errors in quadrature. One
observes that the errors on $\delta R_{\tau}^{kl}$ for all moments are
dominated by the present uncertainty in the strange quark mass, but also
the uncertainties resulting from higher order QCD corrections play some
r\^ole.

Since experimental results for the individual $R$-ratios
$R_{\tau}=3.642 \pm 0.012$ and $R_{\tau,S}=0.1625 \pm 0.0066$ \cite{dy:02}
are available, we can employ our result for the $(0,0)$ moment
\begin{equation}
\label{delR00th}
\delta R_{\tau,th} \;=\; 0.229 \pm 0.030 \,,
\end{equation}
to calculate the value of $|V_{us}|$. Assuming unitarity of the CKM matrix
in order to express the CKM matrix element $|V_{ud}|$ in terms of $|V_{us}|$,
we find
\begin{equation}
\label{VusUT}
|V_{us}| \;=\; 0.2179 \pm 0.0044_{\rm exp} \pm 0.0009_{\rm th} \;=\;
0.2179 \pm 0.0045\,.
\end{equation}
The first given error is the experimental uncertainty due to $R_{\tau}$ and
$R_{\tau,S}$ whereas the second error stems from the theoretical value for
$\delta R_{\tau}$. Our result for $|V_{us}|$ is in good agreement with the
PDG average although at present the uncertainty is roughly twice as large and
it is only one sigma away from $|V_{us}|$ as obtained from the CKM unitarity
fit. However, the current uncertainty is dominated by the experimental
uncertainty in $R_{\tau,S}$, and if this shrinks by a factor of two, also a
corresponding reduction in the uncertainty for $|V_{us}|$ would be reached.
Then the determination of $|V_{us}|$ from the $\tau$ sum rule could reach a
precision better than the current world average. In addition, if the value of
$m_s$ will become better known in the future, there is also room for
improvement on the theory side.

A second possibility to calculate $|V_{us}|$ from $\tau$ decays into strange
particles with net strangeness would be to directly employ the theoretical
prediction for $R_{\tau,S}/|V_{us}^2|$. Estimating the uncertainties in
analogy to table~\ref{tab6}, we obtain
\begin{equation}
\label{Rqus}
R_{\tau,S}/|V_{us}^2| \;=\; 3.395 \pm 0.069 \,.
\end{equation}
Since for $R_{\tau,S}$, contrary to $\delta R_{\tau}$, also the $D=0$
contribution $\delta^{(0)}$ is present, in this case the uncertainty is
dominated by the perturbative QCD error, although also the error on the
strange quark mass matters somewhat. Together with the experimental value
$R_{\tau,S}=0.1625 \pm 0.0066$, again we are in a position to extract
$|V_{us}|$ with the result:
\begin{equation}
\label{VusRus}
|V_{us}| \;=\; 0.2188 \pm 0.0044_{\rm exp} \pm 0.0022_{\rm th} \;=\;
0.2188 \pm 0.0049\,.
\end{equation}
We observe that our findings of eqs.~\eqn{VusUT} and \eqn{VusRus} are
completely compatible. The experimental uncertainty in the latter result is
identical to the one of \eqn{VusUT}, whereas the theory error, due to the
contribution from perturbation theory, is larger. To conclude, when improved
experimental data will be available, we expect that with our first approach a
precise determination of $|V_{us}|$ from $\delta R_{\tau}$ can be achieved.

\newsection{Comparison with other analyses}

Our determination of the strange quark mass from hadronic $\tau$ decays
presented above can readily be compared to the recent analysis
\cite{cdghpp:01}, the most notable difference being our replacement of
the badly behaved longitudinal contributions by phenomenological
parametrisations. In ref.~\cite{cdghpp:01} the final result for the strange
mass was $m_s(M_\tau)=120^{+21}_{-26}$ assuming the value for $|V_{us}|$ as
obtained from a fit to the unitarity of the CKM matrix. This finding is in
perfect agreement to our result \eqn{msUT} in this case, although through
the usage of the phenomenological scalar and pseudoscalar spectral functions,
a reduction of the uncertainties could be achieved in our investigation.

Other analyses of $m_s$ from hadronic $\tau$ decays, which have been published
in the last years, can be found in refs.
\cite{ckp:98,aleph:99,pp:99,kkp:00,km:00,dchpp:01}. A detailed comparison of
these analyses has recently been worked out by Maltman in ref.~\cite{mal:02}.
There it was observed that, once all analyses are updated by using common
values for $R_{\tau,S}$ and of the CKM matrix element $|V_{us}|$, the resulting
strange mass values all cluster around $m_s(2\,\gev)=115\,\mev$ for a unitarity
$|V_{us}|$ and $m_s(2\,\gev)=102\,\mev$ for the PDG average, being in perfect
agreement to our findings of eqs.~\eqn{msUT} and \eqn{msLR} respectively.

Sum rule determinations of the strange quark mass have also been performed on
the basis of the divergence of the vector or axialvector spectral functions
alone \cite{cfnp:97,jam:98,mk:02,jop:02}. Being directly proportional to
$m_s^2$, these sum rules are very sensitive to the strange mass and the
relevant spectral functions are identical to the ones employed to replace
the scalar and pseudoscalar longitudinal contribution to the $\tau$ sum rule.
The outcomes of these analyses were $m_s(2\,\gev)=100\pm 12\,\mev$ for the
pseudoscalar finite energy sum rule \cite{mk:02}, and
$m_s(2\,\gev)=99\pm 16\,\mev$ in the case of the scalar Borel sum rule
\cite{jop:02}. Again, these results are in very good agreement to our value
for $m_s$ of eq.~\eqn{msLR} for the PDG average for $|V_{us}|$, whereas in
the case of the unitarity $|V_{us}|$ the result from the $\tau$ sum rule is
roughly one sigma higher. However, at present, due to the large uncertainties
no further conclusions from this slight discrepancy can be reached.

The status of the extraction of $m_s$ from the hadronic $e^+e^-$ cross
section is less clear. In an analysis which took into account flavour-breaking
differences, analogous to the $\tau$ sum rule, Narison \cite{nar:95} obtained
$m_s(2\,\gev)=144\pm 21\;\mev$. In refs.~\cite{mal:98,mw:99} it was then
pointed out that large isospin breaking corrections significantly lower the
result for the strange mass to about $m_s(2\,\gev)=95\;\mev$ and yield
considerably larger uncertainties of the order of $45\;\mev$. A recent
improved analysis taking into account further SU(3)-breaking differences of
hadronic correlators now yields a value of $m_s(2\,\gev)=129\pm 24\;\mev$
\cite{nar:99}. Nevertheless, in our opinion further work in this channel is
needed, before a definite conclusion can be reached \cite{ejs:03}.

Recent reviews of determinations of the strange quark mass from lattice QCD
have been presented in refs.~\cite{lub:01,gm:01,kan:01,wit:02}, with the
conclusions $m_s(2\,\gev)= 108\pm 15\,\mev$ and $m_s(2\,\gev)= 90\pm 20\;\mev$
in the quenched and unquenched cases respectively. In the unquenched case the
error in these results is dominated by the uncertainty resulting from dynamical
fermions, whereas the calculations of $m_s$ in the quenched theory, based for
example on the kaon mass, are already very precise. However, also in the
quenched lattice simulations additional uncertainties arise due to the fact
that strange masses extracted from different quantities display systematic
deviations \cite{wit:02}. Nevertheless, the agreement between lattice QCD and
QCD sum rule determinations of $m_s$ is already rather satisfactory.

\begin{figure}
\begin{center}
\includegraphics[angle=270, width=12cm]{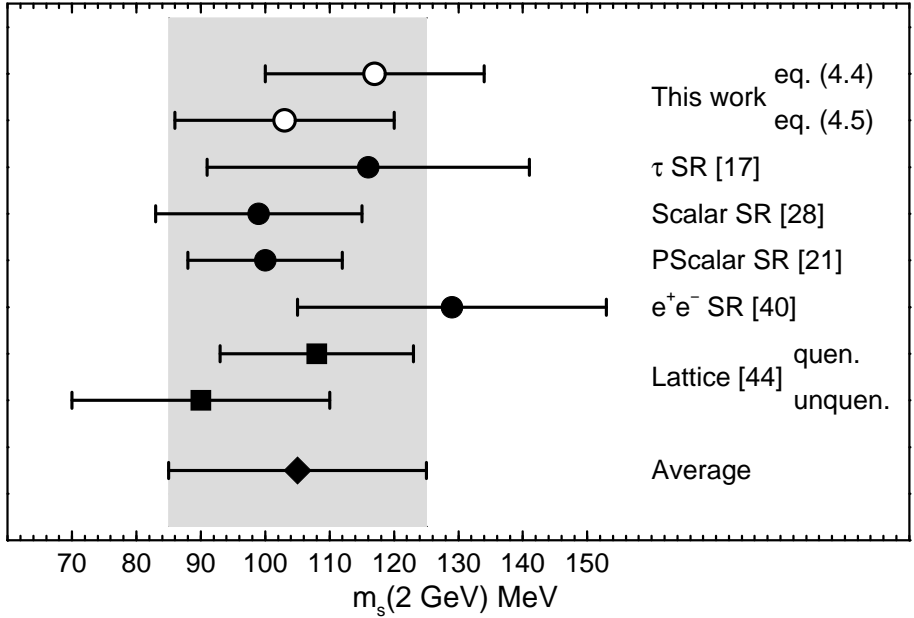}
\end{center}
\caption{Summary of results for $m_s(2\,\gev)$.
\label{fig1}}
\end{figure}
In figure~\ref{fig1}, we display a summary of the status of strange mass
determinations from QCD sum rules and lattice QCD. The two open circles
represent the results of this work given in eqs.~\eqn{msUT} and \eqn{msLR}.
The full circles correspond to other sum rule determinations discussed in this
section, while the full squares are averages for lattice QCD simulations in
the quenched and unquenched case. Finally, the full diamond corresponds to
our average value $m_s(2\,\gev)=105\pm 20\,\mev$ which we have used in
sections~3 and 5 as an input.

\newsection{Conclusion}

The most severe drawback of determinations of the strange quark mass from
the flavour SU(3)-breaking difference \eqn{delRtaukl} so far has been the
bad behaviour of the longitudinal contributions \eqn{RtauklL} within
perturbative QCD. This drawback could be circumvented by replacing the badly
behaved longitudinal contributions with corresponding phenomenological
parametrisations of the relevant spectral functions. To gain confidence in
this treatment, in section~3 we verified that the scalar and pseudoscalar
$\tau$-like finite energy sum rules are satisfied by themselves within the
uncertainties, which however are large for the theoretical expressions.

Already in previous analyses it was found that the extracted strange mass
depends sensitively on the value of the CKM matrix element $|V_{us}|$ which is
used in eq.~\eqn{delRtaukl}. Thus, in our numerical analysis of section~4,
we decided to consider two different cases: in the first one, like in
ref.~\cite{cdghpp:01}, we employed the value $|V_{us}|=0.2225\pm0.0021$ which
results from a fit to the unitarity of the CKM matrix \cite{pdg:02}. At a
scale of $2\,\gev$ our resulting strange mass \eqn{msUT} was found to be
\begin{equation}
m_s(2\,\gev) \;=\; 117 \pm 17 \,\mev \,.
\end{equation}
On the other hand using the PDG central average $|V_{us}|=0.2196\pm0.0026$,
which is based on the analyses \cite{lr:84,cknrt:01,cl:02}, the strange mass
obtained is lowered by roughly one standard deviation and takes the value
\begin{equation}
m_s(2\,\gev) \;=\; 103 \pm 17 \,\mev \,.
\end{equation}
One should, however, remark that these results are obtained from an average
over the different $(k,0)$ moments which have been analysed and that the
individual results for the strange mass display a sizeable dependence on the
number of the moment. Nevertheless, the deviations to our central averages
are within one sigma for all individual moments.

Taking advantage of the strong sensitivity of the flavour-breaking $\tau$
sum rule on $|V_{us}|$, in section~5 we inverted the line of reasoning and
determined the CKM matrix element $|V_{us}|$ from the same sum rule by assuming
an average for the strange quark mass as extracted from other sources. The
result thus obtained is
\begin{equation}
|V_{us}| \;=\; 0.2179 \pm 0.0045\,,
\end{equation}
where the error is largely dominated by the experimental result
$R_{\tau,S}=0.1625 \pm 0.0066$ \cite{dy:02}. A reduction of this uncertainty
by a factor of two would therefore result in a corresponding reduction in the
error on $|V_{us}|$, which would lead to a determination more precise than the
current PDG average. Furthermore, precise experimental measurements of
$R_{\tau,S}^{kl}$ and the SU(3)-breaking differences $\delta R_\tau^{kl}$
would open the possibility to determine both $m_s$ and $|V_{us}|$
simultaneously. This can hopefully be achieved with the BABAR and BELLE
$\tau$ data samples in the near future.

Such an independent access on $|V_{us}|$ would be extremely welcome in view
of the fact that at present the situation about unitarity in the first row of
the quark-mixing matrix is slightly confusing. Whereas with the PDG average
$|V_{us}| = 0.2196 \pm 0.0026$, based solely on the analysis of $K_{e3}$
decays, unitarity would be violated at the two sigma level, preliminary
results of the E865 experiment at Brookhaven National Laboratory \cite{sher:02} 
would correspond to $|V_{us}| = 0.2278 \pm 0.0029$ \cite{cir:02}, thus being
completely compatible with the unitarity of the CKM matrix.\footnote{Larger
values of $|V_{us}|$ from $K_{e3}$ decays have been advocated before in ref.
\cite{fks:00}.} This preliminary value would, however, be about two standard
deviations away from our result for $|V_{us}|$ presented above. In addition,
one should keep in mind that also shifts in the value for $|V_{ud}|$ are
certainly possible. Anyhow, with the upcoming improvements on this subject,
we are confident that the question about unitarity in the CKM matrix can be
resolved in the near future.

\bigskip
\subsection*{Acknowledgements}
E.G. is indebted to MECD (Spain) for a F.P.U. Fellowship. This work has been
supported in part by the European Union RTN Network EURIDICE Grant No.
HPRN-CT2002-00311 (E.G., A.P. and J.P.), by MCYT (Spain) Grants No.
FPA2000-1558 (E.G. and J.P.), and FPA-2001-3031 (A.P. and M.J.), by GVA Grant
No. CTIDIB/2002/24 (A.P.) and by Junta de Andaluc\'{\i}a Grant No. FQM-101
(E.G. and J.P.). M.J. would also like to thank the Deutsche
Forschungsgemeinschaft for support.

\newpage

\end{document}